\documentclass[12pt]{article} 
\usepackage[english]{babel}
\title{ Mirror World versus large extra dimensions}
\author{ Z.~K.~Silagadze 
\vspace*{3mm} \\
Budker Institute of Nuclear Physics,  630 090,
Novosibirsk, Russia }
\date{}

\begin{document}
\large
\maketitle

\begin{abstract}
Recently proposed low scale quantum gravity scenario is expected to have
a significant impact on the Mirror World hypothesis. Some aspects of this 
influence is investigated here, assuming that the fundamental gravity scale
is near a TeV. It is shown that future colliders will be capable of producing 
the mirror matter, but an experimental signature, which will distinguish
such events from the background, is not clear. 
The ``smoking gun'' signals of the Mirror World would be an observation of 
decays like $\Upsilon(2S)\to \tilde \chi_{b2} \gamma $. But unfortunately 
the expected branching ratios are very small.
Finally, it is shown that a mirror supernova will be quite
a spectacular event for our world too, because a considerable amount of 
ordinary energy is expected to be emitted in the first few seconds. 
\end{abstract}

\newpage

\section{Introduction}
Despite firmly established V-A character of weak interactions, ``the violation
of parity invariance is, remarkably, still an open question'' \cite{1}.
It is not yet excluded experimentally that every ordinary particle is 
accompanied by its mirror twin. This dramatic duplication of the world restores
left-right (and time reversal) symmetry of Nature, provided that the mirror 
particles experience V+A (mirror) weak interactions \cite{1,2}. The idea
itself dates back to the seminal paper of Lee and Yang \cite{3}, but it 
remained rather unfruitful until a phenomenological analysis of 
Kobzarev, Okun and Pomeranchuk \cite{4} appeared, where the term ``Mirror 
World'' was coined for the first time. A number of papers had followed
\cite{5}, and it became clear that the main, if not only, connection between
two worlds should be provided by gravity.   

Recently the Mirror World scenario has revived in connection with neutrino
physics \cite{1,2,6,7,8}. Parallel Standard Model \cite{6,7,8} -- the Standard
Model doubled by the mirror sector, appears to be a viable candidate for the
Standard Model extension. The new parity symmetry (which also interchanges
ordinary and mirror particles) can still be violated spontaneously \cite{6}
during electroweak symmetry breaking, leading to a quite different 
macrophysics for the Mirror World \cite{6}. But esthetically more appealing
seems the possibility for the full Lorentz Group to be an unbroken symmetry
of Nature. Although such Exact Parity Model \cite{7} assumes that the 
symmetry breaking patterns are strictly correlated in two (ordinary and 
mirror) worlds -- an ultimate example of the Einstein, Podolsky, and Rosen 
paradox \cite{9}.   

Besides data on neutrino oscillations, which may support 
existence of sterile neutrino(s)
(see however \cite{10}), where are some other phenomena which can be 
also interpreted as indicating towards the Mirror World: 
\begin{itemize}
\item The mirror matter can constitute a considerable fraction of the dark 
matter \cite{11}.
\item The observed gravitational microlensing events can be caused by
mirror stars \cite{8,12}.
\item The deficit of local luminous matter revealed by the recent Hubble 
Space Telescope star counts \cite{13,14} was predicted by Blinnikov and Khlopov
(1983) \cite{5} as a result of mirror stars existence.
\item The mirror neutrinos and mirror stars can play crucial role in cosmic 
Gamma-ray Bursts \cite{14,15}.
\end{itemize}

\noindent Gravity is the main connector between our and mirror worlds. But it
was recently suggested \cite{16} that the apparent weakness of gravity
can be just a low energy phenomenon caused by the existence of new 
sub-millimeter scale spatial dimensions, in which gravitons propagate freely
in contrast to the Standard Model fields which are confined to 
a three-dimensional  wall (``3-brane'').
In this case, at high energies of  about a TeV, gravity becomes strong, 
comparable in strength to the other interactions. Upcoming collider \cite{17}
and/or gravity \cite{18} experiments will soon uncover whether such low scale
quantum gravity scenario has something to do with reality. Meanwhile we can
speculate about the impact that the existence of the large extra dimensions 
will have on communications between our and mirror worlds.   

It may happen that the visible matter and the mirror matter are located in
two different 3-branes \cite{19}. Such literally parallel world will be 
connected to the ordinary one necessarily very weakly due to exchange of
massive Kaluza-Klein excitations \cite{19}. Remarkably enough, recent
alternative proposal of Randall and Sundrum \cite{20} for solving the 
Hierarchy Problem is based on the model with two parallel 3-branes!

But in this work we will assume that both the ordinary and the mirror 
particles live in the same 3-brane. In this case low scale quantum gravity
implies several new phenomenological consequences for mirror matter searches.
 
\section{Mirror matter production at colliders}
One immediate new feature of a TeV scale quantum gravity is the possibility
for mirror particles to be produced at future high energy colliders. Using
Feynman rules given in \cite{21}, we can get, for example (mirror particles
are denoted by tilde throughout the paper)
\begin{equation}
\frac{d\sigma(e^+e^-\to\tilde e^+\tilde e^-)}{dt}=~\frac{\pi s^2}{32\Lambda^8}
\left \{ 1+10\frac{t}{s}+42\left(\frac{t}{s}\right)^2
+64\left(\frac{t}{s}\right)^3+32\left(\frac{t}{s}\right)^4 \right \},
\label{eq1} \end{equation}
\noindent where $\Lambda \sim 1 {\mathrm TeV}$ is an ultraviolet cutoff energy
for the effective low-energy theory, presumably of the order of the bulk Planck
mass \cite{21}. For the total cross-section we obtain
\begin{equation}
\sigma(e^+e^-\to\tilde e^+\tilde e^-)=\frac{\pi s^3}{160\Lambda^8}\approx
7.6\left (\frac{s}{{\mathrm TeV^2}}\right )^3
\left(\frac{{\mathrm TeV}}{\Lambda}\right)^8~{\mathrm pb}.
\label{eq2} \end{equation}

\noindent Other examples are
\begin{eqnarray} & &
\frac{d\sigma(e^+e^-\to\tilde \gamma \tilde \gamma)}{dt}=~\frac{\pi
s^2}{8\Lambda^8}\left (-\frac{t}{s}\right )\left (1+\frac{t}{s}\right )
\left (1+2\frac{t}{s} +2\frac{t^2}{s^2} \right ), \nonumber \\ & &
\frac{d\sigma(\gamma \gamma \to\tilde \gamma \tilde \gamma)}{dt}=~
\frac{\pi s^2}{2\Lambda^8}\left [ \left( 1+\frac{t}{s}\right )^4 +
\frac{t^4}{s^4}\right ]. \label{eq3} \end{eqnarray}
\noindent Which translates into the following total cross sections
($\frac{1}{2}$ is included to account for identity of the final state photons)
\begin{eqnarray} & &
\sigma(e^+e^-\to\tilde \gamma \tilde \gamma)=
\frac{\pi s^3}{320\Lambda^8}\approx
3.8\left (\frac{s}{{\mathrm TeV}^2}\right)^3
\left (\frac{{\mathrm TeV}}{\Lambda}\right )^8~{\mathrm pb}, \nonumber \\ & & 
\sigma(\gamma \gamma \to\tilde \gamma \tilde \gamma)=
\frac{\pi s^3}{20\Lambda^8}\approx
61\left (\frac{s}{{\mathrm TeV}^2}\right )^3
\left (\frac{{\mathrm TeV}}{\Lambda}\right )^8~{\mathrm pb}.
\label{eq4} \end{eqnarray}
\noindent
Note that the above given estimates, being obtained from the effective theory,
are valid only for energies lower than the cutoff scale $\Lambda$.

As we see, future colliders may have sizeable ability for the mirror matter
production. To experimentalists regret, there is no useful signature for such
kind of reactions, and a quest for them looks like hunting for the black cat
in a dark room. May be more clear signature have reactions accompanied by
the initial-state radiation \cite{22}. But we suspect severe background 
problems here, in particular from the real graviton emission \cite{23}. 
As for the mirror matter two-photon production, in the equivalent photon
approximation \cite{24} we have, for example
\begin{equation}
\sigma(e^+e^-\to e^+e^-\tilde\gamma\tilde\gamma)=
\frac{\alpha^2}{\pi^2}\int\limits_0^1 \frac{dz}{z}\left [ f(z)(L-1)^2+
\frac{1}{3}\ln^3{z}\right ] \sigma_{\gamma\gamma\to\tilde\gamma\tilde\gamma}
(zs).
\label{eq5} \end{equation}
\noindent Where $L=\ln{\frac{s}{m_e^2}}$ and
$$
f(z)=\left (1+\frac{1}{2}z\right )^2\ln{\frac{1}{z}}-\frac{1}{2}(1-z)(3+z).
$$
\noindent Substituting $\sigma_{\gamma\gamma\to\tilde\gamma\tilde\gamma}$
from (\ref{eq4}) and integrating, we get
\begin{equation}
\sigma(e^+e^-\to e^+e^-\tilde\gamma\tilde\gamma)=
\frac{\alpha^2 s^3}{180\pi\Lambda^8}\left [\frac{121}{400}(L-1)^2-
\frac{2}{3} \right ].
\label{eq6} \end{equation}
\noindent For $s\approx 1{\mathrm TeV}^2$ this is about $10^4$ times smaller
than $\sigma(\gamma\gamma\to\tilde\gamma\tilde\gamma)$ at the same center of
mass energy. Besides, considerable background is expected in this channel too.
For example, from the real graviton two-photon production \cite{25}.

\section{Quarkonium -- mirror quarkonium oscillations}
Virtual graviton exchange 

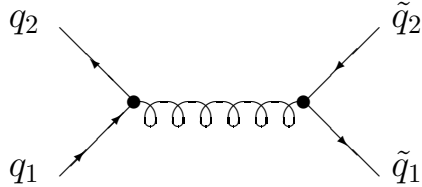
\input FEYNMAN
\begin{center}
\begin{picture}(20000,9000)
\drawline\fermion[\SE\REG](2000,8000)[4000]
\global\Xone=\pbackx
\global\Yone=\pbacky
\drawarrow[\NW\ATBASE](\pmidx,\pmidy)
\put(\pbackx,\pbacky){\circle*{500}}
\global\advance\pfrontx by -1900
\put(\pfrontx,\pfronty){$q_2$}
\drawline\fermion[\SW\REG](\pbackx,\pbacky)[2000]
\drawarrow[\NE\ATBASE](\pmidx,\pmidy)
\drawline\gluon[\E\REG](\Xone,\Yone)[6]
\startphantom
\drawline\fermion[\SW\REG](\pfrontx,\pfronty)[2000]
\stopphantom
\drawline\fermion[\SW\REG](\pbackx,\pbacky)[2000]
\global\advance\pbackx by -1900
\put(\pbackx,\pbacky){$q_1$}
\drawarrow[\NE\ATBASE](\pmidx,\pmidy)
\drawline\fermion[\NE\REG](\gluonbackx,\gluonbacky)[4000]
\drawarrow[\SW\ATBASE](\pmidx,\pmidy)
\global\advance\pbackx by 500
\put(\pbackx,\pbacky){$\tilde q_2$}
\put(\pfrontx,\pfronty){\circle*{500}}
\drawline\fermion[\SE\REG](\gluonbackx,\gluonbacky)[4000]
\drawarrow[\SE\ATBASE](\pmidx,\pmidy)
\global\advance\pbackx by 500
\put(\pbackx,\pbacky){$\tilde q_1$}
\end{picture}
\end{center}
\noindent leads to the ordinary--mirror matter mixing due
to the following low energy effective 4-fermion interaction \cite{21}
\begin{eqnarray} & & \hspace*{30mm}
H= 8\pi C_4 m_q m_{\tilde q} \frac{n+1}{n+2}\bar \psi \psi \bar {\tilde \psi}
 \tilde \psi - \nonumber \\ & & 
-\frac{\pi}{2}C_4\left [ (q_1+q_2)\cdot (\tilde q_1+\tilde q_2)
\bar \psi \gamma^\mu \psi \bar {\tilde \psi} \gamma_\mu \tilde \psi +
\bar \psi (\hat {\tilde q_1}+\hat {\tilde q_2})\psi 
\bar {\tilde \psi} (\hat q_1+\hat q_2) \tilde \psi \right ], 
\label{eq7} \end{eqnarray}
\noindent where $n$ is the number of the large extra dimensions and in
numerical estimates we will assume
\begin{equation}
C_4\sim \frac{1}{\Lambda^4}, \; \; \; \Lambda \sim 1{\mathrm TeV}.
\label{eq8} \end{equation}

As a result of this mixing, heavy C-even quarkonia can oscillate into their
mirror counterparts, and hence disappear from our world \cite{26}. 
The same effect takes place, of course, for the light quarkonia and 
positronium also, but apparently miniscule magnitudes makes it completely 
irrelevant in these cases.

Having at hand (\ref{eq7}) and using that in the weak binding 
nonrelativistic limit the state vector of quarkonium can be represented
as a superposition of the free quark-antiquark states \cite{27}, it is
straightforward to calculate quarkonium -- mirror quarkonium transition
amplitude. For tensor quarkonium (like $\chi_{b2}$) and for its mirror 
analog we get the following mass matrix
$$ \left ( \begin{array}{cc} M_\chi & -\epsilon M_\chi \\
-\epsilon M_\chi & M_\chi \end{array} \right ), $$
\noindent with
\begin{equation}
\epsilon = \frac{6N_c C_4}{M_\chi}\dot{R}^2(0).
\label{eq9} \end{equation}
\noindent Where $M_\chi$ is the quarkonium mass and $\dot{R}(0)$ is the
derivative at $r=0$ of its radial wave function. $N_c=3$ accounts for quark 
color.

States with definite $M_\chi(1-\epsilon)$ and $M_\chi(1+\epsilon)$ masses are
$$\chi_+=\frac{1}{\sqrt{2}}(\chi+\tilde \chi), \; \; \; 
\chi_-=\frac{1}{\sqrt{2}}(\chi-\tilde \chi). $$
\noindent Therefore, the probability for ordinary quarkonium, with lifetime
$\tau_\chi$, to oscillate into the mirror quarkonium and disappear is
\begin{equation}
{\mathrm Br}(\chi\to \tilde \chi)=\int\limits_0^\infty e^{-\Gamma_\chi t}
\sin^2{(\epsilon M_\chi t)}~\frac{dt}{\tau_\chi}=\frac{2(\epsilon M_\chi)^2}
{\Gamma_\chi^2+4\epsilon^2 M_\chi^2}\approx 2\left (\frac{\epsilon M_\chi}
{\Gamma_\chi}\right )^2.
\label{eq10} \end{equation}
\noindent Let us estimate this invisible branching ratio for $\chi_{b2}$,
for example. The derivative at $r=0$ of the radial wave function depends
on the potential used, but for our purposes the following numbers look 
realistic \cite{28}
$$\dot{R}(0)\approx 1.4~GeV^5 \; \; \; {\mathrm and} \; \; \;
\Gamma_{\chi_{b2}}\approx 200~keV.$$
\noindent Then
$$\frac{\epsilon M_\chi}{\Gamma_\chi}=6N_C\frac{\dot{R}^2(0)}
{\Lambda^4 \Gamma_\chi}\approx 1.3\cdot 10^{-7} $$
\noindent and
\begin{equation}
Br(\chi_{b2}\to \tilde \chi_{b2}) \approx 3\cdot 10^{-14}.
\label{eq11} \end{equation} 

The ``smoking gun'' signal for the mirror world existence would be an 
observation of $\Upsilon(2S)\to \tilde \chi_{b2} \gamma $ decay, which will 
have very clear signature. The probability for such kind of decay can be 
estimated to be about $2\cdot 10^{-15}$, from the equation (\ref{eq11}) and
from the known branching ratio for $\Upsilon(2S)\to \chi_{b2} \gamma $ 
decay. Unfortunately this seems to be too small to be of practical interest.
 
\section{Ordinary luminosity from mirror supernova}
As we have seen, mirror matter production in electron--positron collisions
can have sizeable magnitude. As a result, some part of a mirror supernova 
energy will be released in our world too. Let us estimate this effect, taking
into account only $\tilde e^+ \tilde e^-\to e^+ e^-,\; \gamma \gamma $ 
reactions with the total cross section
\begin{equation}
\sigma=\frac{3\pi s^3}{320 \Lambda^8}. 
\label{eq12} \end{equation}
\noindent There are also some other channels for the mirror--ordinary energy
transfer (see, for example, analogous considerations for axion emission rates
\cite{29}, and for graviton emission rates \cite{30}). But our purpose here 
is just to show that the effect may be significant in principle. So, to 
estimate the mirror -- ordinary energy transfer, we will use the cross section
(\ref{eq12}).

The ordinary energy emissivity per unit volume per unit time of a mirror
supernova core with a temperature $T$ is given by the thermal average over the
Fermi-Dirac distribution \cite{29,30}.
\begin{equation}
\dot{q}=\int dn_+~dn_-\frac{(E_+ +E_-)}{E_+E_-}\{E_+E_-
\sqrt{(\vec{v}_+-\vec{v}_-)^2+(\vec{v}_+\cdot\vec{v}_-)^2-v_+^2v_-^2} 
\}\sigma,
\label{eq13} \end{equation}
\noindent where
\begin{equation}
dn_\pm=\frac{2d\vec{q}_\pm}{(2\pi)^3}\left [ \exp{\left (\frac{E_\pm
\mp\mu_e}
{T}\right )} + 1 \right ]^{-1},
\label{eq14} \end{equation}
\noindent $\mu_e$ and $-\mu_e$ being the chemical potentials for (mirror)
electrons and \linebreak positrons.

The flux factor $\{E_+E_-
\sqrt{(\vec{v}_+-\vec{v}_-)^2+(\vec{v}_+\cdot\vec{v}_-)^2-v_+^2v_-^2} 
\}$
is separated \cite{31} because 
it is Lorentz invariant and can be  easily calculated in the center of mass
frame as being $\frac{1}{2}s$, if the electron mass is neglected. In the 
supernova frame
$$ s=2E_+E_-(1-\cos{\theta_{+-}}). $$
\noindent
Now it is straightforward to perform integrations in (\ref{eq13}) and we get
\begin{equation}
\dot{q}=\frac{6T^{13}}{25\pi^3\Lambda^8}[I_5(\nu)I_6(-\nu)+
I_5(-\nu)I_6(\nu)],
\label{eq15} \end{equation}
\noindent where 
$$ \nu=\frac{\mu_e}{T} \; \; \; {\mathrm and} \; \; \;
I_n(\nu)=\int\limits_0^\infty dx \frac{x^n}{\exp{(x+\nu)}+1}. $$

\noindent Let us compare (\ref{eq15}) to the neutrino emissivity by 
supernova \cite{32} (only the leading term is shown)
\begin{equation}
\dot{q}_{\nu \bar \nu}=\frac{2G_F^2T^{9}}{9\pi^5}(C_V^2+C_A^2)
[I_3(\nu)I_4(-\nu)+I_3(-\nu)I_4(\nu)],
\label{eq16} \end{equation}
\noindent where $C_A=\frac{1}{2}, \; C_V=\frac{1}{2}+2\sin^2{\Theta_W}$ and
$G_F$ is the Fermi coupling constant.

\noindent From (\ref{eq15}) and (\ref{eq16}) we get
\begin{equation}
\frac{\dot{q}}{\dot{q}_{\nu \bar \nu}}=\frac{27\pi^2}{25(C_V^2+C_A^2)}~
\frac{T^4}{\Lambda^8 G_F^2}~\frac{I_5(\nu)I_6(-\nu)+
I_5(-\nu)I_6(\nu)}{I_3(\nu)I_4(-\nu)+I_3(-\nu)I_4(\nu)}.
\label{eq17} \end{equation}
\noindent For a moderate temperature $T=30{\mathrm MeV}$, chemical potential
$\mu_e\approx 345{\mathrm MeV}$ \cite{30}
and $\Lambda \sim 1{\mathrm TeV}$, the last equation gives
\begin{equation}
\frac{\dot{q}}{\dot{q}_{\nu \bar \nu}}\approx 1.4\cdot 10^{-16}.
\label{eq18} \end{equation}
\noindent But in the first $\sim 10$ seconds the neutrino luminosity from 
a supernova is tremendous \cite{33}: $L_{\nu \bar \nu}\approx 3\cdot 10^{45}
W$ for each species of neutrino. Therefore a small number (\ref{eq18})
translates into the following ordinary luminosity of a mirror supernova
$L\approx 4 \cdot 10^{29} W$. This is about $10^3$ times larger than the
solar luminosity! 

\section{conclusions}
To summarize, if the fundamental scale for quantum gravity is about a TeV and
if the Mirror World exists in the same 3-brane where our world resides, we 
will have new possibilities to probe the Mirror World. Even the near future
colliders will have an ability to produce mirror particles. But an 
experimental signature of such events is unclear. It seems that the Mirror
World will not be immediately discovered due to this effect after the large 
extra dimensions will become an experimental fact (if such an exciting event 
really happens in future experiments), but the discovery  will wait for the 
detailed theory of quantum  gravity and precise experiments.

A clear signal of the Mirror World would be an observation of the 
$\Upsilon(2S)\to \tilde \chi_{b2} \gamma $ decay. But the expected 
branching ratio $\sim 10^{-15}$ is too small to be observable.  

The most drastic impact we have for mirror supernovas. For some 10
seconds they can give a flash in our world brighter than thousand  suns!
I think it is interesting to look for such events.

Finally, let us note that the effects considered in this paper are caused
by the virtual graviton exchange. Therefore they depend only weakly on the 
number of the large extra dimensions.


\end{document}